
\input harvmac
\lref\GutzToda{M.~Gutzwiller, Ann.\ Phys.\ {\bf 133} (1981) 304--331.}
\lref\KomGCh{I.~V.~Komarov, Theor.\ Math.\ Phys.\ {\bf 50} (1982) 265--270.}
\lref\QISM{L.~D.~Faddeev and L.~A.~Takhtajan, Sov.\ Sci.\ Rev.\ Math.\
{\bf C1} (1981) 107. \hfill\break
V.~E.~Korepin, N.~M.~Bogoliubov and A.~G.~Izergin,
{\it ``Quantum Inverse Scattering Method and Correlation Functions'',}
Cambridge Univ.\ Press (1993).}
\lref\Arn{V.~I.~Arnold, {\it ``Mathematical Methods of Classical Mechanics''},
   2nd edition, Springer (1989).}
\lref\BabVi{O.~Babelon and C-M.~Viallet, Phys.\ Lett.\ {\bf B237} (1990)
411--416.}
\lref\nonSLN{B.~A.~Dubrovin, J.\ Sov.\ Math.\ {\bf 28} (1985) 20--50.
 \hfill\break
  A.~I.~Bobenko, A.~G.~Reyman and M.~A.~Semenov-Tian-Shansky,
Commun.\ Math.\ Phys.\ {\bf 122} (1989) 321-354. }
\lref\BelDr{A.~A.~Belavin and V.~G.~Drinfeld, Funct.\ Anal.\ Appl.\ {\bf 16}
(1982) 159--180; Funct.\ Anal.\ Appl.\
{\bf 17} (1983) 220--221; Sov.\ Sci.\ Rev.\ {\bf C4} (1984) 93--.}
\lref\FeiFr{B.~Feigin and E.~Frenkel, ``Integrals of motion and quantum
groups'', Preprint YITP/K-1036 (1993), to appear in Proceedings of the
Summer School {\it Integrable Systems and Quantum Groups,} Montecatini
Terme, Italy, June 1993, Lect.\ Notes in Math., Springer.}
\lref\FFRGaudin{B.~Feigin, E.~Frenkel and N.~Reshetikhin, Commun.\
Math.\ Phys.\ {\bf 166} (1994) 27--62.}
\lref\FeiFr{B.~Feigin and E.~Frenkel, ``Integrals of motion and quantum
groups'', Preprint YITP/K-1036 (1993), to appear in Proceedings of the
Summer School {\it Integrable Systems and Quantum Groups,} Montecatini
Terme, Italy, June 1993, Lect.\ Notes in Math., Springer.}
\lref\Miwa{B.~Davies, O.~Foda, M.~Jimbo, T.~Miwa and A.~Nakayashiki,
 Commun.\ Math.\ Phys.\ {\bf 151} (1993) 89--153.}
\lref\Mumford{D.~Mumford, {\it ``Tata lectures on Theta. I''}, Birkh\"auser
 (1983).}
\lref\HarAdHur{ M.~R.~Adams, J.~Harnad and J. Hurtubise, Commun.\ Math.\
Phys.\ {\bf 155} (1993) 385--413.}
\lref\Scott{D.~R.~D.~Scott, J.\ Math.\ Phys.\ {\bf 35} (1994) 5831--5843.
\hfill\break
M.I.~Gekhtman, Commun.\ Math.\ Phys.\ {\bf 167} (1995) 593--605.}
\lref\OlPer{M.~A.~Olshanetsky and A.~M.~Perelomov, Phys.\ Rep.\
 {\bf 71} (1981) 313-400; {\bf 94} (1983) 313--404.}
\lref\Macd{I.~G.~Macdonald, {\it ``Symmetric functions and Hall polynomials''},
Clarendon press (1979).}
\lref\BabFlu{H.~M.~Babujian and R.~Flume, Mod.\ Phys.\ Lett.\ {\bf A9}
(1994) 2029--2040.}
\lref\ReshVar{N.~Reshetikhin and A.~Varchenko, ``Quasiclassical asymptotics
of solutions of the KZ equations'', preprint {\sf hep-th/9402126.}}
\lref\FFR{B.~Feigin, E.~Frenkel and N.~Reshetikhin, ``Gaudin model,
Bethe ansatz and correlation functions at the critical level'', preprint
{\sf hep-th/9402022.}}

\lref\alggeo{I.~M.~Krichever and S.~P.~Novikov, Russ.\ Math.\ Surv.\
 {\bf 35}:6 (1980) 53--79. \hfill\break
 A.~B.~Dubrovin, Russ.\ Math.\ Surv.\ {\bf 36}:2 (1981) 11--92.}
\lref\seminal{H.~Flaschka and D.~W.~McLaughlin, Progr.\ Theor.\ Phys.\
{\bf 55} (1976) 438--456. \hfill\break
I.~M.~Gel'fand and L.~A.~Diki\u{\i}, Funct.\ Anal.\ Appl.\
{\bf 13} (1979) 8--20. \hfill\break
S.~P.~Novikov and A.~P.~Veselov, Proc.\ Steklov Inst.\ Math.\
{\bf 3} (1985) 53--65.}
\lref\QCD{L.~N.~Lipatov, JETP Lett.\ {\bf 59} (1994) 596--599.
\hfill\break
L.~D.~Faddeev and G.~P.~Korchemsky, Phys.\ lett.\ {\bf B342} (1995) 311--322.}
\lref\GCh{I.~V.~Komarov and V.~V.Zalipaev, J.\ Phys.\ A: Math.\ Gen.\
{\bf 17} (1984) 31--49. \hfill\break
 I.~V.~Komarov and \Kuz, J.\ Sov.\ Math.\ {\bf 47} (1989)
2459--2466. \hfill\break
 \Kuz\ and A.~V.~Tsiganov, J.\ of Physics A: Math.\ Gen.\
{\bf 22} (1989) L73--L79. \hfill\break
 I.~V.~Komarov and E.~I.~Novikov, Phys.\ Lett.\ {\bf A186} (1994)
 396--402.}

\lref\rmCM{\EKS, ``Dynamical $r$-matrices for the elliptic Calogero-Moser
model'', Preprint LPTHE-93-42 (1993); {\sf hep-th/9308060};
Algebra i Analiz {\bf 6} (1994) 227--237 (in Russian). \hfill\break
H.~W.Braden and T.~Suzuki, Lett.\ Math.\ Phys.\ {\bf 30} (1994) 147--158.
   \hfill\break
 H.~W.Braden and T.~Suzuki, Phys.\ Lett.\ {\bf A192} (1994) 17--21.}
\lref\BiAvBab{E.~Billey, J.~Avan and O.~Babelon,
Phys.\ Lett.\ {\bf A186} (1994) 114--118.}
\lref\EiEn{J.~C.~Eilbeck, V.~Z.~Enol'skii, V.~B.~Kuznetsov and
D.~V.~Leykin, Phys.\ Lett.\ {\bf A180} (1993) 208--214. \hfill\break
 J.~C.~Eilbeck, V.~Z.~Enol'skii, V.~B.~Kuznetsov and A.~V.~Tsiganov,
 J.\ Phys.\ A: Math.\ Gen.\ {\bf 27} (1994) 567--578. \hfill\break
 J.~C.~Eilbeck, V.~Z.~Enol'skii, V.~B.~Kuznetsov and
D.~V.~Leykin, ``Linear $r$-matrix algebra for a hierarchy of
one-dimensional particle systems separable in parabolic coordinates'',
Sfb 288 Preprint No. 110, Differentialgeometrie und
Quantenphysik, Berlin, 28pp., February 1994.}
\lref\Brz{T.~Brzezi\'nski, Phys.\ Lett.\ {\bf B325} (1994) 401--408.}

\lref\KuzGaudin{\Kuz, J.\ Math.\ Phys.\ {\bf 33} (1992) 3240--3254.
  \hfill\break
  E.~G.~Kalnins, \Kuz\ and W.~Miller, Jr., J.\ Math.\ Phys.\ {\bf 35} (1994)
1710--1731.}
\lref\KuzNeumann{\Kuz, Funkts.\ Anal.\ Appl.\ {\bf 26} (1992) 302-304.}
\lref\BabTal{O.~Babelon and M.~Talon, Nucl.\ Phys.\ {\bf B379} (1992)
  321--329.}
\lref\KomCoulomb{I.~V.~Komarov, J.\ Phys.\ A: Math.\ Gen.\ {\bf 21}
   (1988) 1191--1198.}
\lref\KomKuzEuler{I.~V.~Komarov and \Kuz, J.\ Phys.\ A: Math.\ Gen.\
{\bf 24} (1991) L737--L742.}
\lref\MacF{A.~J.~MacFarlane, Nucl.\ Phys.\ {\bf B386} (1992) 453--467.}
\lref\BrMcf{T.~Brzezi\'nski and A.~J.~Macfarlane, J.\ Math.\ Phys.\
{\bf 35} (1994) 3261--3276.} 
\lref\HarWin{J.~Harnad and P.~Winternitz, ``Classical and quantum integrable
systems in $\tilde{gl}(2)^{+*}$ and separation of variables'',
Preprint CRM-1921 (1993); {\sf hep-th/9312035}. \hfill\break
 J.~Harnad and P.~Winternitz, Lett.\ Math.\ Phys.\ {\bf 33}
(1995) 61--74. \hfill\break
 M.~A.~del Olmo, M.~A.~Rodr\'{\i}guez and P.~Winternitz,
``The conformal group $SU(2,2)$ and integrable systems on a Lorentzian
hyperboloid'', Preprint CRM-2194 (1994); {\sf hep-th/9407080}.}

\lref\EnKuzSal{V.~Z.~Enol'skii, \Kuz\ and M.~Salerno, Physica D {\bf 68}
(1993) 138--152.}
\lref\ChJoKuz{P.~L.~Christiansen, M.~F.~J{\o}rgensen and V.~B.~Kuznetsov,
Lett.\ Math.\ Phys.\ {\bf 29} (1993) 165--173.}
\lref\KuzTsigToda{\Kuz\ and A.~V.~Tsiganov, ``Separation of variables for the
quantum relativistic Toda lattices'', Research Report 94-07, University
of Amsterdam, Amsterdam, pp.1-13, 1994; {\sf hep-th/9402111}.}
\lref\KuzTsigBC{\Kuz\ and A.~V.~Tsiganov, Zapiski Nauchn.\ Semin.\ LOMI
 {\bf 172} (1989) 88--98.}

\lref\KomKuzKowal{I.~V.~Komarov and \Kuz, Theoret.\ Math.\  Phys.\ {\bf 73}
(1987) 1255--1263.}

\lref\SklGoryachev{\EKS, J.\ Soviet Math.\ {\bf 31} (1985), 3417--3431.}
\lref\SklToda{\EKS, ``The quantum Toda chain'', in {\it  ``Non-linear
      equations in classical and quantum field  theory''}, ed.\ N.~Sanchez.
      (Lecture Notes in Physics, v.\ 226), Springer, 1985, p.\ 196--233.}
\lref\SklXYZ{\EKS, J.\ Soviet Math.\ {\bf 46} (1989), 1664--1683.}
\lref\SklBoundary{\EKS, J.\ Phys.\ A: Math.\ Gen.\ {\bf 21} (1988) 2375--2389.}
\lref\SklXXZ{\EKS, J.\ Soviet Math.\ {\bf 46} (1989), 2104--2111.}
\lref\SklGaudin{\EKS, J.\ Soviet Math.\ {\bf 47} (1989), 2473--2488.}
\lref\SklNLS{\EKS, J.\ Sov.\ Math.\ {\bf 19} (1982) 1546--1595.}
\lref\SklNLSiv{\EKS, J.\ Phys.\ A: Math.\ Gen.\ {\bf 22} (1989) 3551--3560.
  \hfill\break
  \EKS, Leningrad Math.\ J.\ {\bf 1} (1990) 515--534.}
\lref\SklSinhG{\EKS, Nucl.\ Phys.\ {\bf B326} (1989) 719--736.}
\lref\SklFBA{\rm \EKS, ``Functional Bethe ansatz'', in {\it ``Integrable
      and superintegrable systems''}, ed.\ B.~A.~Kupershmidt (World
      Scientific, 1990), p.\ 8--33.}
\lref\SklNankai{\EKS, ``Quantum Inverse Scattering Method. Selected
      Topics'', in {\it ``Quantum Group and Quantum Integrable Systems''}
      (Nankai Lectures in Mathematical Physics), ed.\ Mo-Lin Ge
      (World Scientific, 1992), p.\ 63--97.}
\lref\SklSLthreeCl{\EKS, Commun.\ Math.\ Phys.\ {\bf 150} (1992), 181--191.}
\lref\SklSLthreeQ{\EKS, ``Separation of variables in the quantum
     integrable models related to the Yangian ${\cal Y}[sl(3)]$'',
     Newton Inst.\ Preprint NI-92013 (1992); {\sf hep-th/9212076};
      Zap.\ nauchn.\ semin.\ POMI {\bf 205} (1993) 166--178.}
\lref\SklKuz{\Kuz\ and \EKS, ``Separation of variables in $A_2$ type
     Jack polynomials'', University of Tokyo Preprint UTMS 95-10 (1995).}
\font\sc=cmcsc10
\font\sf=cmss10

\font\bbd=msbm10
\font\frak=eufm10
%

\let\tilde=\widetilde

\def\tr{{\rm tr}}

\def\res{\mathop{\rm res}\limits}
\def\one#1{#1^{\raise5pt\hbox{$\scriptstyle\!\!\!\!1$}}\,{}}
\def\two#1{#1^{\raise5pt\hbox{$\scriptstyle\!\!\!\!2$}}\,{}}
\def\three#1{#1^{\raise5pt\hbox{$\scriptstyle\!\!\!\!3$}}\,{}}
\def\phi{\varphi}
\def\eps{\varepsilon}
\def\a{\alpha}

\def\d{\delta}
\def\D{\Delta}
\def\l{\lambda}
\def\L{\Lambda}
\def\s{\sigma}
\def\ket#1{\left|#1\right>}

\def\vac{\ket{0}}
\def\comment#1{}

\def\id{\hbox{{1}\kern-.25em\hbox{l}}}
\def\EKS{E.~K.~Sklyanin}
\def\Kuz{V.~B.~Kuznetsov}
\def\half{{1 \over 2}}
\def\frac#1#2{{#1 \over #2}}
\def\strutdepth{\dp\strutbox}
\def\specialstar{\vtop to \strutdepth{
   \baselineskip\strutdepth
   \vss\llap{* }\null}}
\def\marginalstar{\strut\vadjust{\kern-\strutdepth\specialstar}}
\def\add#1{\marginalstar{\sf $\lceil$#1$\rfloor$}}
\def\?{\add{?}}
\def\CC{\hbox{\bbd C}}
\def\RR{\hbox{\bbd R}}
\def\r{\hbox{\frak r}}
\def\t{\hbox{\frak t}}
\def\K{{\cal K}}
\def\LL{\hbox{\frak L}}
\def\ZZ{\hbox{\bbd Z}}
\def\dd{\partial}
\def\SoV{\hbox{So}\kern-0.10em\hbox{V}}

\def\num{\the\secno.\the\meqno}
\def\onum{(\num)}
\def\numadd{\the\secno.\the\meqno
             \global\advance\meqno by1}

\rightline{\sf UTMS 95-9}
\rightline{\sf solv-int/9504001}
\Title{}{\vbox{\centerline{Separation of Variables. New Trends.}}}
\centerline{Evgueni\ K.\ \sc Sklyanin}
\bigskip
\centerline{\it  Department of Mathematical Sciences, Tokyo University,
 Tokyo, Japan}
\centerline{\it and}
\centerline{\it Steklov Mathematical Institute,
St.Petersburg, Russia}
\vskip 1.0cm

Talk given at the International Workshop
 ``Quantum Field Theory, Integrable Models and Beyond'',
YITP, Kyoto, 14--17 Feb, 1994;
to be published in Progr.\ Theor.\ Phys.\ Suppl.\ No.\ 118 (1995) 35--60.

\vskip 1cm

\centerline{\bf Abstract}
The review is based on the author's papers since 1985 in which a new approach
to the separation of variables (\SoV) has being developed.
It is argued that \SoV, understood
generally enough, could be the most universal tool to solve
integrable models of the classical and quantum mechanics.
It is shown that the standard construction of the action-angle
variables from the poles of the Baker-Akhiezer function can be
interpreted as a variant of \SoV, and moreover, for many particular
models it has a direct quantum counterpart. The list of the models
discussed includes XXX and XYZ magnets, Gaudin model,
Nonlinear Schr\"odinger  equation, $SL(3)$-invariant
magnetic chain.
New results for the 3-particle quantum Calogero-Moser system are
reported.

\vfill
\Date{}

\centerline{\bf Contents}
\noindent \S1. Introduction \hfill\break
\noindent \S2. Separation of variables: general notions \hfill\break
\indent 2.1 Basic definitions \hfill\break
\indent 2.2 Magic recipe \hfill\break
\indent 2.3 $r$-matrix formalism \hfill\break
\noindent \S3. $GL(N)$-type models \hfill\break
\indent 3.1 Classical case \hfill\break
\indent 3.2 Quantization \hfill\break
\noindent \S4. XXX magnetic chain \hfill\break
\noindent \S5. Infinite volume limit \hfill\break
\noindent \S6. Classical XYZ magnet \hfill\break
\noindent \S7. 3-particle Calogero-Moser model \hfill\break
\noindent \S8. Discussion \hfill\break
\noindent References

\newsec{Introduction}

The separation of variables (\SoV), at least, in its most elementary forms
such as \SoV\ in cartesian, spherical or ellipsoidal coordinates,
is an indispensable part of the basic mathematical/physical curriculum.
Briefly, \SoV\ can be characterized as a reduction
of a multidimensional problem to a set of one-dimensional ones.
Originated from the works of D'Alembert and Fourier (wave theory)
and Jacobi (Hamiltonian mechanics),
the \SoV\ for the long time was the only known method of ``exact''
solution of problems of mathematical physics.
However, in the last decades the new
techniques, including Inverse Scattering Method (ISM) as well as its
quantum version (QISM) together with Bethe Ansatz,
seemed to oust the \SoV\ out of fashion.

The aim of the present review is to draw attention to the recent progress
in understanding \SoV\ and its relations to ISM and QISM. I am going to
argue that \SoV\ is far yet from
being outdated and, even more, has good chances to remain as
the most universal method of solving completely integrable (classical
and quantum) models.
There are two basic observations which give support to that claim.
First: for the classical integrable systems subject
to ISM the standard construction of the action-angle variables using
the poles of the Baker-Akhiezer function is in fact equivalent to
a separation of variables. And second: in many cases it is possible to
find the precise quantum analog of this construction.

This point of view has being gradually clarified since my
first publications \refs{\SklGoryachev, \SklToda} of 1985
which were deeply influenced by
I.~V.~Komarov (St.~Petersburg State University) who used to insist that
\SoV\ and ISM are closely related and to underline that the notion of \SoV\
should not be restricted to the purely coordinate change of variables
as it was frequently done. Generally speaking, \SoV\ can be produced by a
complicated canonical transformation involving both coordinates and
momenta. In the quantum case the canonical transformation should be
replaced with a unitary operator. The papers of Gutzwiller on 3,4-particle
Toda lattice \refs{\GutzToda} and of Komarov on
Goryachev-Chaplygin top \refs{\KomGCh}  had provided the first examples of how
such a unitary operator could be guessed using known classical \SoV.
In \refs{\SklGoryachev, \SklToda} the results of Komarov and Gutzwiller
were reproduced using the machinery of QISM \QISM.
The algebraic techniques of QISM (R-matrix method) opened the way to
methodical construction of quantum \SoV\ for the whole classes of integrable
models generated by different R-matrices (solutions to the Yang-Baxter
equation). Though the  realization of this program is far yet from
completion, the list of the models, to which the new approach has been
applied successfully, grows steadily and presently includes:
 \item{$\bullet$} XXX magnetic chain \refs{\SklFBA,\SklNankai}
 \item{$\bullet$} XXX Gaudin model \SklGaudin, see also
\refs{\KuzGaudin\KuzNeumann\BabTal\KomCoulomb\KomKuzEuler
\HarWin\MacF{--}\BrMcf}
 \item{$\bullet$} XYZ magnetic chain (classical case) \SklXYZ
 \item{$\bullet$} Goryachev-Chaplygin top \refs{\KomGCh,\SklGoryachev},
see also \GCh\
 \item{$\bullet$} Toda lattice \SklToda, including relativistic case
\KuzTsigToda\ and boundary conditions \KuzTsigBC; see also \ChJoKuz.
 \item{$\bullet$} Nonlinear Schr\"odinger equation (infinite volume)
\refs{\SklNLSiv}
 \item{$\bullet$} sinh-Gordon model (infinite volume) \SklSinhG
 \item{$\bullet$} $SL(3)$ magnetic chain \refs{\SklSLthreeCl, \SklSLthreeQ}
 \item{$\bullet$} 3-particle Calogero-Moser model, see Section 7 of the
present paper

The early stages of the work were summarized in the reviews
\refs{\SklFBA, \SklNankai},
where the term ``Functional Bethe Ansatz'' was used instead \SoV, the latter
one seeming now to be the more accurate.
I tried to avoid in the present review the detailed discussions of
particular models which one can find in \refs{\SklFBA, \SklNankai}
but, instead, to summarize the development of the field using the
unifying concept of the Baker-Akhiezer function. The importance of choosing
a proper normalization of B-A function is stressed.
More attention is paid to the particular models which were not discussed
in the previous reviews (Nonlinear Schr\"odinger equation in the infinite
volume, classical XYZ magnet). New results are reported concerning \SoV\
in the quantum 3-particle Calogero-Moser model (elliptic and trigonometric).

\newsec{\SoV: general notions}
\subsec{Basic definitions}
Let us start with the classical case. Consider a Hamiltonian mechanical
system having a finite number $D$ of degrees of freedom and integrable in the
Liouville's sense. It means \Arn\ that one is given a $2D$-dimensional
symplectic manifold (phase space) and $D$ independent hamiltonians $H_j$
commuting with respect to the Poisson bracket
\eqn\commHcl{
 \{H_j,H_k\}=0\qquad j,k=1,\ldots,D.
}

A system of canonical variables $(\vec p,\vec x)$
\eqn\canpb{
 \{x_j,x_k\}=\{p_j,p_k\}=0,\qquad  \{p_j,x_k\}=\d_{jk}
}
will be called {\it separated} if there exist $D$ relations of the form
\eqn\sovcl{
\Phi_j(x_j,p_j,H_1,H_2,\ldots,H_D)=0\qquad j=1,2,\ldots,D
}
binding together each pair $(p_j, x_j)$ and the hamiltonians $H_n$.
Note that fixing the values of hamiltonians $H_n=h_n$ one obtains from
\sovcl\ an explicit factorization of the Liouville tori into the
one-dimensional ovals given by equations
\eqn\tori{
\Phi_j(x_j,p_j,h_1,h_2,\ldots,h_D)=0\qquad j=1,2,\ldots,D.
}

The {\it action} function $S(\vec h,\vec x)$ is defined \Arn\ as the
generating function of the canonical transformation from $(\vec p,\vec x)$
to the {\it action-angle} variables $(\vec I,\vec \phi)$ that is
\eqn\actang{
  \vec h=\vec h(\vec I), \qquad
  p_j=\frac{\dd S(\vec h(\vec I),\vec x)}{\dd x_j}, \qquad
  \phi_j=\frac{\dd S(\vec h(\vec I),\vec x)}{\dd I_j}.
}
and satisfies the Hamilton-Jacobi equation
\eqn\HJeq{
  H_j\left(\frac{\dd S(\vec h(\vec I),\vec x)}{\dd\vec x},\vec x\right)
     =h_j(\vec I)
}
for each $H_j$ considered as a function of $(\vec p,\vec x)$.
The relations \tori\ allow immediately to split the complete solution to
{\it partial differential equation} \HJeq\ into the sum of terms
\eqn\Ssplit{
    S(\vec h,x_1,\ldots,x_D)=S_1(\vec h,x_1)+\cdots+S_D(\vec h,x_D)
}
satisfying each the {\it ordinary differential equation}
\eqn\sepeqcl{
 \Phi_j(x_j,\dd_{x_j}S_j,h_1,h_2,\ldots,h_D)=0
}
(we consider $\vec h$ in $S_j(x_j)\equiv S_j(\vec h,x_j)$ as fixed parameters).

The last result justifies apparently the term ``separation of variables''.
It is important to warn the reader that many authors
restrict the notion of \SoV\ only with the situations
when the separated variables $(\vec p,\vec x)$ are obtained from
some original canonical variables, say $(\vec P,\vec Q)$, by purely
{\it coordinate} transform
\eqn\cotan{
    x_j=x_j(\vec Q), \qquad
    p_j=\sum_k \frac{\dd Q_k}{\dd x_j}P_k.
}

That condition, though reasonable for a certain class of problems,
would be, however, too restrictive for our purposes excluding almost all the
examples we are going to consider.

Let us examine now the quantum case. The condition \commHcl\ is replaced by
the commutativity of quantum operators
\eqn\commHq{
 [H_j,H_k]=0\qquad j,k=1,\ldots,D
}
the relations \sovcl\ retaining the same form
\eqn\sovq{
    \Phi_j(x_j,p_j,H_1,H_2,\ldots,H_D)=0
}
with the only difference that now we have to fix some ordering of
the noncommutative operators $x_j$, $p_j$, $H_n$. Let us assume that
the operators in \sovq\ are ordered exactly as they are enlisted that is
$x$ always precedes $p$ and $p$ precedes $H$ (the relative ordering of $H_n$
is of no importance since they are commutative \commHq).

It is convenient to work in the $x$-representation that is to realize
the quantum states as the functions $\Psi(\vec x)$ from some Hilbert space
${\cal H}$.
The canonical operators $x_j$ and $p_j$
\eqn\CCR{
   [x_j,x_k]=[p_j,p_k]=0\qquad  [p_j,x_k]=-i\hbar\d_{jk}
}
can be realized respectively as the multiplication and differentiation
$p_j=-i\hbar\dd_{x_j}$ operators.

Let $\Psi$ be a common eigenfunction
\eqn\HPsi{ H_j\Psi=h_j\Psi }
of the commuting
hamiltonians. Then, under assumptions made about the operator ordering
and the realization of $(\vec p,\vec x)$ it follows from \sovq\ that
$\Psi(\vec x)$ satisfies the equations
\eqn\sepeqPsi{
    \Phi_j(x_j,-i\hbar\dd_{x_j},h_1,h_2,\ldots,h_D)\Psi=0
}
which suggests immediately the factorization of $\Psi$
\eqn\wffact{
    \Psi(x_1,\ldots,x_D)=\prod_{j=1}^D \psi_j(x_j)
}
into functions $\psi_j(x_j)$ of one variable only satisfying
the equation
\eqn\sepeqpsi{
    \Phi_j(x_j,-i\hbar\dd_{x_j},h_1,h_2,\ldots,h_D)\psi_j(x_j)=0.
}

In complete analogy with the classical case, the original multidimensional
spectral problem \HPsi\ is reduced to the set  of the one-dimensional
multiparameter spectral problems \sepeqpsi. For determining the admissible
values of $D$ spectral parameters $h_j$ one has thus the system of $D$
equations \sepeqpsi\ supplemented with appropriate boundary conditions
determined by the Hilbert space ${\cal H}$.
The function $\Phi_j$ in \sepeqPsi\ can be thought of as
a symbol of a pseudodifferential operator. In particular, \sepeqpsi\
becomes an ordinary differential equation if $\Phi_j$ is a polynomial in $p$
or a finite-difference equation if $\Phi_j$ is a trigonometric polynomial
in $p$.

Semiclassicaly,
$$ \Psi(\vec x)=e^{\frac{i}{\hbar}S(\vec x)},\qquad
    \psi_j(x_j)=e^{\frac{i}{\hbar}S_j(x_j)} $$
which provides the correspondence of the formulas \Ssplit\ and \wffact.

\subsec{Magic recipe}
Let us return now to the classical case and discuss
the question how to find a \SoV\ for a given
integrable system. In the XIXth and the beginning
of the present century for a number of models of classical mechanics,
 such as Neumann model or various cases of the rigid body motion,
the \SoV\ was found by guess or some more or less {\it ad hoc} methods.

Here we shall discuss the latest and seemingly the most powerful method
based on the {Baker-Akhiezer function}.
Suppose, as it is always done in the Inverse Scattering Method (ISM),
that our commutative hamiltonians $H_j$ can be obtained as the spectral
invariants of some matrix $L(u)$
of dimensions $N\!\!\times\!\! N$, called $L$ or {\it Lax operator}, whose
elements are functions on the phase space and depend also on an additional
parameter $u$ called {\it spectral parameter}. It means that $H_j$ can be
expressed in terms of the coefficients $t_n(u)$ of the {\it characteristic
polynomial} $W(z,u)$ of the matrix $L(u)$
\eqn\charpol{\eqalign{
& W(z,u)=\det(z-L(u))=\sum_{n=0}^N (-1)^n t_n(u) z^{N-n}, \cr
& t_0(u)=1,   \qquad t_n(u)=\tr\bigwedge^n L(u), \qquad t_N(u)=\det L(u).
}}

The {\it characteristic equation}
\eqn\spcurve{
  W(z,u)=0
}
defines the eigenvalue $z(u)$ of $L(u)$ as a function on the corresponding
$N$-sheeted Riemannian surface of parameter $u$.
The {\it Baker-Akhiezer function} $\Omega(u)$ is defined then as the
eigenvector of $L(u)$
\eqn\eigvL{
   L(u)\Omega(u)=z(u)\Omega(u)
}
corresponding to the eigenvalue $z(u)$.

Since an eigenvector is defined up to a scalar factor, to exclude the ambiguity
in the definition of $\Omega(u)$ one has to {\it fix a normalization}
of $\Omega(u)$ imposing a linear constraint
\eqn\normOm{
       \sum_{n=1}^N \a_n(u)\Omega_n(u)=1
}
where $\a_n(u)$ may, generally speaking, depend also on the dynamical
variables. A normalization being fixed, $\Omega(u)$ becomes
a meromorphic function on the Riemannian surface \spcurve.

The poles $x_j$ of the Baker-Akhiezer function play an important role in
ISM \alggeo. In particular, the time
evolution of $x_j$ for the hamiltonian flow with any of the commuting
hamiltonians $H_n$ can be expressed explicitely in terms of the
Riemannian theta-functions corresponding to the spectral curve \spcurve.
Moreover, it was observed that for many models the variables $x_j$
Poisson commute and, together with the corresponding eigenvalues
$z_j\equiv z(x_j)$ of $L(x_j)$, or some functions $p_j$ of $z_j$,
provide a set of separated canonical variables
for the Hamiltonians $H_n$. It should be mentioned that, though the
seminal papers \seminal\ contain all the
necessary results,
the possibility of their interpretation in terms of \SoV\ was not recognized
at that time.

One of the reasons why the poles of $\Omega(u)$ could provide a \SoV\
is easy to understand. Since $z_j=z(x_j)$ is an eigenvalue of $L(x_j)$
the pair $(z_j,x_j)$ should lie on the spectral curve \spcurve
\eqn\chareq{
        W(z_j,x_j)=0.
}

It remains to observe that, if $z_j$ is a function of $p_j$,
the equation \chareq\ fits exactly the form
\sovcl\ since the
coefficients $t_n(x_j)$ of the characteristic polynomial \charpol\
contain nothing except $x_j$ and the hamiltonians $H_k$.

However, the relations \sovcl\ alone are not enough to produce \SoV.
It is necessary that, in addition, the number of the poles $x_j$ be
exactly the number of degrees of freedom $D$ and that
the variables $(p_j(z_j),x_j)$ be
canonical \canpb. Those properties are by no means obvious and the last one
is rather difficult to verify.
To perform the calculation, it is necessary to transform the above
definitions of $(z_j,x_j)$ into  a more convenient form.
Let $\Omega^{(j)}=\res_{u=x_j}\Omega(u)$. From \eigvL\ and \normOm\
there follow, respectively, the eigenvalue equation and the normalization
condition for $\Omega^{(j)}$
\eqn\Omj{
     L(x_j)\Omega^{(j)}=z_j\Omega^{(j)}, \qquad
     \sum_{n=1}^N\a_n(x_j)\Omega^{(j)}_n=0.
}

Let $\vec\a(u)$ denote the one-row matrix $(\a_1(u),\ldots,\a_N(u))$.
The existence of a nonzero solution $\Omega^{(j)}$ to the problem \Omj\
is equivalent to the condition
\eqn\rankAL{
     {\rm rank}\pmatrix{\vec\a(x_j) \cr L(x_j)-z_j\id}=N-1
}
which, in turn, can be expressed generically as vanishing of any of two
minors of order $N$, for instance
\edef\bigminors{\onum}
$$ \eqalignno{
&   \left|\matrix{
   \a_1(x) & \a_2(x) & \a_3(x) & \ldots & \a_N(x) \cr
   L_{21}(x)& L_{22}(x)-z & L_{23}(x) & \ldots & L_{2N}(x) \cr
   L_{31}(x)& L_{32}(x) & L_{33}(x)-z & \ldots & L_{3N}(x) \cr
  \ldots & \ldots & \ldots & \ldots & \ldots \cr
   L_{N1}(x) & L_{N2}(x) & L_{N3}(x) & \ldots & L_{NN}(x)-z }\right|=0,
&(\num{\rm a})\cr
&    \left|\matrix{
   \a_1(x) & \a_2(x) & \a_3(x) & \ldots & \a_N(x) \cr
   L_{11}(x)-z& L_{12}(x) & L_{13}(x) & \ldots & L_{1N}(x) \cr
   L_{31}(x)& L_{32}(x) & L_{33}(x)-z & \ldots & L_{3N}(x) \cr
  \ldots & \ldots & \ldots & \ldots & \ldots \cr
   L_{N1}(x) & L_{N2}(x) & L_{N3}(x) & \ldots & L_{NN}(x)-z }\right|=0.
&(\num{\rm b})
}$$
\advance\meqno by1

The pairs $(z_j,x_j)$ are obtained then as the roots of the system
of equations \bigminors\ which allows, in principle, to count them and to
calculate the
Poisson brackets between them.

In the sections devoted to the nonlinear Schr\"odinger equation and
XYZ model we shall need the formulas for the $N=2$ case.
The equations \bigminors\ become then quite simple:
\eqn\aazx{
  \left|\matrix{\a_1(x) & \a_2(x) \cr
                L_{21}(x) & L_{22}(x)-z}\right|=0, \qquad
  \left|\matrix{\a_1(x) & \a_2(x) \cr
                L_{11}(x)-z & L_{12}(x)}\right|=0.
}

Eliminating $z$ one obtains the equation for $x_j$
\eqn\defBaa{
   B(x_j)=0, \qquad
   B=\a_1^2 L_{12}-\a_1\a_2(L_{11}-L_{22})
         -\a_2^2 L_{21}.
}

The eigenvalue $z_j$ is obtained then from
\eqn\defzaa{
     z_j=\left(L_{11}-\frac{\a_1}{\a_2}L_{12}\right)_{u=x_j}
        =\left(L_{22}-\frac{\a_2}{\a_1}L_{21}\right)_{u=x_j}
}
(for the sake of brevity we have omitted the argument $u$ in $L(u)$ and
$\a(u)$ in the last two formulas).

Generally speaking, there is no guarantee that
one obtains from \bigminors\ the canonical P.b.\ \canpb\ for some $p_j(z_j)$.
Amazingly, it turns out to be true for
a fairly large class of integrable models, though the fundamental
reasons responsible for such effectiveness of the magic recipe:
{\it ``Take the poles of the properly normalized Baker-Akhiezer function
and the corresponding
eigenvalues of the Lax operator and you obtain a \SoV''}, are still
unclear.  The key words in the above recipe are ``the properly normalized''.
The choice of the proper normalization $\vec\a(u)$ of $\Omega(u)$ can be quite
nontrivial (see below the discussion of the XYZ magnet) and for some
integrable models the problem remains unsolved \refs{\nonSLN}.

\subsec{$r$-matrix formalism}
Given a particular $L$ operator and a normalization of $\Omega$,
one is able, in principle,
to calculate from \bigminors\ the Poisson brackets for $(z_j,x_j)$ though
it could be a
formidable task. There are, however, techniques which simplify the
calculation and allow to verify \SoV\ for whole families of $L$ operators
instead of handling them individually.

According to a remarkable theorem proved by Babelon and Viallet \BabVi,
the commutativity of the spectral invariants $t_n(u)$ \charpol\ of the matrix
$L(u)$
\eqn\commt{
  \{t_m(u),t_n(v)\}=0
}
is equivalent to the existence of a matrix $r_{12}(u,v)$ of order
$N^2\times N^2$ such that the Poisson brackets
between the components of $L$ are represented in the commutator form
\eqn\pbl{
   \{\one L(u),\two L(v)\}=[r_{12}(u,v),\one L(u)]-[r_{21}(v,u),\two L(v)]
}
where the standard notation is introduced \SklNankai:
$ \one L\equiv L\otimes\id$,\
$\two L\equiv \id\otimes L,$\
$r_{21}(u,v)={\cal P}r_{12}(u,v){\cal P},$
and ${\cal P}$ is the permutation operator:
${\cal P}x\otimes y =y\otimes x$,\ $\forall x,y\in\CC^N$.

Generally speaking, the matrix $r_{12}(u,v)$ is a function of dynamical
variables. So far, very little is known of such $r$-matrices, apart of
few particular examples \refs{\rmCM\BiAvBab\EiEn{--}\Brz}.
The best studied one is the case
of purely numeric ($c$-number) $r$-matrices satisfying the {\it classical
Yang-Baxter equation}
\eqn\cYBE{
 [r_{12}(u_1,u_2),r_{13}(u_1,u_3)+r_{23}(u_2,u_3)]
-[r_{13}(u_1,u_3),r_{32}(u_3,u_2)]=0
}
which ensures the Jacobi identity for the Poisson bracket \pbl,
and especially, the case of {\it unitary} numeric $r$-matrices,
satisfying, in addition, the relation
\eqn\unitr{
     r_{12}(u_1,u_2)=-r_{21}(u_2,u_1)
}
and depending on the difference $u_1-u_2$.
For such $r$-matrices the relation \pbl\ takes the form
\eqn\pblu{
   \{\one L(u),\two L(v)\}=[r_{12}(u-v),\one L(u)+\two L(v)]
}
and \cYBE, respectively,
\eqn\cYBEu{
 [r_{12}(u),r_{13}(u+v)]+[r_{12}(u),r_{23}(v)]+[r_{13}(u+v),r_{23}(v)]=0.
}

To a unitary numeric $r$-matrix one can associate not only the Poisson
algebra \pblu\ whose right hand side is linear in $L$ but also
the algebra
\eqn\pblquadr{
    \{\one L(u),\two L(v)\}=[r_{12}(u-v),\one L(u)\two L(v)]
}
with the quadratic r.h.s. Formally, \pblquadr\ can be put into the form
\pbl\ with the {\it dynamic} $r$-matrix
$\tilde r_{12}(u,v)=\half(r_{12}(u-v)\two L(v)+\two L(v)r_{12}(u-v))$,\
$\tilde r_{21}(u,v)=-\half(r_{12}(u-v)\one L(u)+\one L(u)r_{12}(u-v))$,
but the very special
structure of $\tilde r_{12}$ allows to consider the formula \pblquadr\
rather as
a modification of \pblu. Obviously, \pblu\ can be obtained from \pblquadr\
if one substitutes $L:=1+\eps L+O(\eps^2)$, $r:=\eps r$
and let $\eps\rightarrow 0$.

Another example of the quadratic P.b.\ algebra associated to a unitary
numeric $r$-matrix is the algebra \SklBoundary
\eqn\rLrL{
   \{\one L(u),\two L(v)\}=[r_{12}(u-v),\one L(u)\two L(v)]
     +\one L(u)r_{12}(u+v)\two L(v)-\two L(v)r_{12}(u+v)\one L(u).
}

There exists a profound algebraic theory of the unitary numeric $r$-matrices
\refs{\BelDr} which allows to classify $r$-matrices in families labelled
by Lie algebras.
A particularly important
example is given for any semisimple Lie algebra ${\cal G}$ by the formula
\eqn\canonr{
     r_{12}(u)=\frac{\rho}{u}\sum_\a I_\a\otimes I_\a
}
where $\rho$ is a numeric constant and $I_\a\in{\cal G}$ is an orthonormal
basis with respect to the Killing
form. The result does not depend on the choice of the basis.
Taking then various finite-dimensional
representations of ${\cal G}$ for the generators $I_\a$ one can obtain
from \canonr\ the family of $r$-matrices related to ${\cal G}$.

In particular, for ${\cal G}=gl(N)$ and the fundamental vector representation,
one has
\eqn\rgln{
   r_{12}(u)=\frac{\rho}{u}{\cal P}.
}

The last example deserves special attention since, so far, it is the
only series of $r$-matrices for which a general \SoV\ construction
is obtained.

\newsec{$GL(N)$-type models}
\subsec{Classical case.}
It turns out that in case of the $GL(N)$-invariant $r$-matrix \rgln\
the normalization of the Baker-Akhiezer function $\Omega(u)$
corresponding to any constant numeric vector $\vec \a$ in \normOm\
produces \SoV. The simplest choice of $\vec\a$ is
\eqn\normGLN{
    \a_1(u)=\ldots=\a_{N-1}(u)=0, \quad \a_N=1
    \qquad \Longleftrightarrow \qquad
   \Omega_N(u)=1, \quad \Omega_j=0.
}

The corresponding separated coordinates $x_j$ are defined as the poles
of $\Omega(u)$, and the canonically conjugated momenta \CCR\ are
$p_j=-\rho^{-1}z_j$ for the linear P.b.\ \pblu\ and $p_j=-\rho^{-1}\ln z_j$
for the quadratic P.b.\ \pblquadr. For the linear P.b.\ case the above
results
were obtained in \refs{\HarAdHur}. The quadratic P.b.\ case was studied in
\refs{\SklGoryachev,\SklSLthreeCl} for $N=2,3$ and generalised to
arbitrary $N$ in  \Scott.

In case of the normalization \normGLN\ the equations \bigminors\ for $(z,x)$
simplify a bit
\edef\twominors{\onum}
$$ \eqalignno{
&   \left|\matrix{
   L_{21}(x)& L_{22}(x)-z & L_{23}(x) & \ldots & L_{2,N-1}(x)\cr
   L_{31}(x)& L_{32}(x) & L_{33}(x)-z & \ldots & L_{3,N-1}(x) \cr
  \ldots & \ldots & \ldots & \ldots & \ldots \cr
   L_{N-1,1}(x) & L_{N-1,2}(x) & L_{N-1,3}(x) & \ldots & L_{N,N-1}(x)-z \cr
   L_{N1}(x) & L_{N2}(x) & L_{N3}(x) & \ldots & L_{N,N-1}(x) }\right|=0,
&(\num{\rm a})\cr
&    \left|\matrix{
   L_{11}(x)-z& L_{12}(x) & L_{13}(x) & \ldots & L_{1,N-1}(x)\cr
   L_{31}(x)& L_{32}(x) & L_{33}(x)-z & \ldots & L_{3,N-1}(x) \cr
  \ldots & \ldots & \ldots & \ldots & \ldots \cr
   L_{N-1,1}(x) & L_{N-1,2}(x) & L_{N-1,3}(x) & \ldots & L_{N,N-1}(x)-z \cr
   L_{N1}(x) & L_{N2}(x) & L_{N3}(x) & \ldots & L_{N,N-1}(x) }\right|=0,
&(\num{\rm b})
}$$
\advance\meqno by1

The equations \twominors\ themselves can be used already
for calculation
of P.b.\ between $x_j$ and $z_j$, see \refs{\HarAdHur,\Scott}.
However, they are not convenient for quantization because of the operator
ordering problem. So, we take one more step and eliminate $z_j$ from
\twominors. The result is one equation
\eqn\Bxj{
     B(x_j)=0
}
for $x_j$ where $B(u)$ is a certain polynomial of degree $N(N-1)/2$
in components of $L(u)$. The corresponding eigenvalue $z_j$ is obtained then
as  the value
\eqn\Axj{
              z_j=A(x_j)
}
of certain function $A(u)$ expressed rationally in components of $L(u)$.

For instance, for $N=2$
\eqn\ABtwo{
       B(u)=L_{21}(u), \qquad A(u)=L_{11}(u).
}

Note that for $u=x_j$ the matrix  $L(u)$ becomes triangular
\eqn\triangL{
 L(x_j)=\pmatrix{
                  z_j & L_{12}(x_j) \cr
                   0  & L_{22}(x_j) }
}
which explains why  its eigenvalue $z_j$ is given by $A(x_j)$.

For $N=3$ one obtains
\eqn\Bthree{
          B(u)=L_{31}(u)\left|\matrix{
               L_{11} & L_{12} \cr
               L_{31} & L_{32}
                   }\right|(u)+
        L_{32}(u)\left|\matrix{
               L_{21} & L_{22} \cr
               L_{31} & L_{32}
                   }\right|(u).
}

There are two possible ways to choose $A(u)$
\eqn\Athree{
 A_1(u)=\frac{\left|\matrix{
               L_{11} & L_{12} \cr
               L_{31} & L_{32}
                   }\right|(u)}{L_{32}(u)},
 \qquad
A_2(u)=-\frac{\left|\matrix{
               L_{21} & L_{22} \cr
               L_{31} & L_{32}
                   }\right|(u)}{L_{31}(u)}
}
which are equivalent modulo $B(x_j)=0\;\Longleftrightarrow\;A_1(x_j)=A_2(x_j)$
\eqn\eqAA{
     z_j=A_1(x_j)=A_2(x_j).
}

The expressions for $A(u)$ and $B(u)$ for the general $N$ are given in \Scott.
We have to warn the reader that due to a different choice of normalization
of $\Omega(u)$ the formulas \ABtwo, \Bthree, \Athree\ differ from those
in \refs{\SklNankai,\SklSLthreeQ,\Scott}.

The equations \Bxj\ and \Axj, like \twominors, can also be used for
calculating the P.b. between $x_j$ and $z_j$. Their advantage for the sake
of quantization is that $B(u)$ Poisson commute
\eqn\commBcl{
   \{B(u),B(v)\}=0
}
which entrains immediately the commutativity of $x_j$
(see the next subsection).

The correct P.b.\ between $x_j$ and $z_j$ are not enough to establish \SoV.
The last condition (usually, easy to verify) is the correct number of
variables $x_j$ which should be equal to the number $D$ of degrees of freedom.
In some degenerate cases, the number of $x_j$ could be less then $D$ and
some additional variables should be added (see example of Calogero-Moser model
in section 7).

To conclude this subsection, let us stress that so far no generalization
is known of the above results to the $r$-matrices corresponding to the
Lie algebras other then the $A_n$ series. The difficulty is that the
simplest normalization \normGLN\ does not work more: the function $B(u)$
has too many zeroes (more than the number $D$ of degrees of freedom)
and they do not commute \nonSLN. Hopefully, some other normalization
\normOm\ will work which remains a challenging problem.

\subsec{Quantization}
In the quantum case the components $L_{mn}$ of the $N\!\times\!N$
matrix $L(u)$ become quantum operators, and the Poisson brackets
should be replaced by some commutation relations sutisfying the
correspondence principle $[\; , \;]=-i\hbar\{\; ,\;\}$.
A nice feature of the Poisson algebras \pblu\ and \pblquadr\ is that
(in contrast with the general case \pbl\ of dynamical $r$-matrices)
it is well known how to quantize them.

In the linear case \pblu\ the quantization is straightforward
\eqn\rLL{
   [\one L(u),\two L(v)]=[\r_{12}(u-v),\one L(u)+\two L(v)],
  \qquad \r(u)=-i\hbar r(u),
}
in the quadratic case \pblquadr\ it is more tricky.
The algebra \pblquadr\ is replaced by
\eqn\RLL{
   R_{12}(u-v)\one L(u)\two L(v)=\two L(v)\one L(u)R_{12}(u-v)
}
where $R(u)$ satisfies the {\it quantum Yang-Baxter equation}
\eqn\QYBE{
  R_{12}(u)R_{13}(u+v)R_{23}(v)=R_{23}(v)R_{13}(u+v)R_{12}(u)
}
and is related to $r(u)$ through the semiclassical expansion
\eqn\semiclR{
       R(u)=1+i\hbar r(u)+O(\hbar^2).
}

In the $GL(N)$ case \rgln\ the quantum $\r$ and $R$ matrices are
\eqn\RGLN{
     \r(u)=-\eta\frac{{\cal P}}{u}, \quad
     R(u)=1+\eta\frac{{\cal P}}{u}, \qquad \eta=i\hbar\rho.
}

The relations \RLL\ define the associative algebra $Y[gl(N)]$ called
yangian of $gl(N)$.

The quantum integrals of motions $t_n(u)$ for the yangian are obtained by
appropriate deformation of the classical formulas \charpol
\eqn\qtn{
 t_n(u)=\tr L(u)\wedge L(u+\eta)\wedge\ldots\wedge L(u+(n-1)\eta)
}
\eqn\commqtn{
    [t_m(u),t_n(u)]=0
}

The quantity $t_N(u)$ ({\it the quantum determinant} of $L(u)$) produces
central elements
\eqn\commtN{
     [t_N(u),L(v)]=0
}
of the yangian. Naturally, on the irreducible representations
$t_N(u)$ is a number-valued function.

The quantum analog of the construction of the \SoV\ based on the
equations \Bxj\ and \Axj\ is found presently only for $N=2$ and $N=3$.
The formulas given below are taken from \refs{\SklNankai,\SklSLthreeQ}
up to small variations due to the different choice of normalization
of $\Omega(u)$.
In the $GL(2)$ case the operator-valued functions $A(u)$ and $B(u)$ are
defined by the same formulas \ABtwo\ as in the classical case.
By virtue
of \RLL\ the operator family $B(u)$ turns out to be commutative
\eqn\commBq{
  [B(u),B(v)]=0
}
which allows to define the operators $x_n$ as the commuting ``operator roots''
of the equation \Bxj\ (for the mathematical details see \SklNankai).
To give a sense to the formula \Axj\ in the quantum case, it is necessary
to fix the operator ordering. Assume that $x$'s in \Axj\ are ordered
to the left that is
\eqn\orderA{
   A(x_j)=\sum_k x_j^k A_k \qquad {\rm for}\qquad
   A(u)=\sum_k u^k A_k.
}

Then, using the commutation relations \RLL\ it is possible to verify
the relations
\eqn\commzx{
 [x_j,x_k]=[z_j,z_k]=0, \qquad z_jx_k=(x_k+\eta\d_{jk})z_j
}
which suggest the realization of the operators $z_j$ as the shift operators in
an appropriate Hilbert space ${\cal H}\ni\Psi(\vec x)$ of functions on
the common spectrum of the operators $x_j$:
\eqn\realz{
  z_j\Psi(\vec x)=\zeta(x_j)\Psi(\ldots,x_j+\eta,\ldots).
}

The choice of the function $\zeta(x)$ in \realz\ is dictated by the
properties of the Hilbert space ${\cal H}$ depending on the concrete
model (see examples below). Note that there is certain liberty in
choosing $\zeta(x_j)$ due to the canonical transformations
\eqn\gauge{
\Psi(\vec x)\rightarrow \prod_j\omega(x_j)\Psi(\vec x)
  \qquad \Longleftrightarrow \qquad
\zeta(x_j)\rightarrow\frac{\omega(x_j+\eta)}{\omega(x_j)}\zeta(x_j).
}

The \SoV\ follows then from the relations
\eqn\qchartwo{
   z_j^2-z_jt_1(x_j)+t_2(x_j)=0
}
which generalize the classical characteristic equation \charpol\ and fit
the form \sovq\ required for the quantum \SoV\ (note the operator
ordering!). Denoting by $\tau_n(u)$ the eigenvalues of the commuting
operators $t_n(u)$ one obtains for the corresponding separated
equation \sepeqpsi\ the finite-difference equation of order 2:
$$
    \zeta(x_j+\eta)\zeta(x_j)\psi_j(x_j+2\eta)
-\zeta(x_j)\tau_1(x_j+\eta)\psi_j(x_j+\eta)+
   \tau_2(x_j)\psi_j(x_j)=0
$$
or in more symmetric form
\eqn\sepeqtwo{
\tau_1(x_j)\psi_j(x_j)=
   \D_-(x_j)\psi_j(x_j-\eta)+\D_+(x_j)\psi_j(x_j+\eta)=0
}
where
$$ \D_+(u)=\zeta(u), \qquad \D_-(u)=\frac{t_2(u-\eta)}{\zeta(u-\eta)}.
$$

In the $GL(3)$ case the quantum $B(u)$ and $A(u)$ are obtained as deformations
of the classical formulas \Bthree\ and \Athree:
\eqn\BthreeQ{
 \eqalign{
B(u)&=L_{31}(u-\eta)[L_{32}(u)L_{11}(u-\eta)-L_{31}(u)L_{12}(u-\eta)] \cr
    &+L_{32}(u-\eta)[L_{32}(u)L_{21}(u-\eta)-L_{31}(u)L_{22}(u-\eta)],
}}
\eqn\AthreeQ{\eqalign{
A_1(u)&=
L_{32}^{-1}(u)[L_{32}(u)L_{11}(u-\eta)-L_{31}(u)L_{12}(u-\eta))], \cr
A_2(u)&=
-L_{31}^{-1}(u)[L_{32}(u)L_{21}(u-\eta)-L_{31}(u)L_{22}(u-\eta))], \cr
}}
\eqn\xAA{
     z_j=A_1(x_j)=A_2(x_j).
}

The rest is similar to $GL(2)$ case.
The quantum characteristic equation is
\eqn\qchareqthree{
    z_j^3-z_j^2t(x_j)+z_j t_2(x_j)-t_3(x_j)=0,
}
and the separated equation is now a
third-order finite-difference equation \SklSLthreeQ.

There is little doubt that the above constructions can be generalized
to the arbitrary values of $N$, though the complicated structure of
$B(u)$ \Scott\ prevents rapid progress.

The analogous results for the linear  commutation relations
\rLL\ can be obtained from the formulas for the quadratic
relations \RLL\ in the limit $L:=1+\eps L+O(\eps^2)$, $\eta:=\eps\eta$,
$\eps\rightarrow 0$. However, though it is clear enough that the
expansion in $\eps$ of the formulas \qtn\ for the quantum integrals
of motion for the yangian should produce, in principle, some commuting
hamiltonians for the algebra \rLL, obtaining explicit formulas for
them when $N$ is arbitrary remains still unsolved problem, to say
nothing about expressions for $A(u)$ and $B(u)$. In case of the $\r$
matrix of the form \canonr\ Feigin and Frenkel \refs{\FeiFr} has
proved for any Lie algebra ${\cal G}$ that the quantum commuting
operators do exist which are deformations of the  spectral
invariants of the classical $L$ matrix though their method of proof does not
produce any effective formulas. Nevertheless, the simplest integrals
of motion are easy to produce. Note that $\t_1(u)=\tr L(u)$ is a trivial
central element of the algebra \rLL, so it can be safely put to be 0,
which corresponds to considering $sl(N)$ instead of $gl(N)$.
The first nontrivial invariant is quadratic in $L(u)$ and coincides
with the classical expression
\eqn\quadrinv{
   \t_2(u)=\half\tr L^2(u), \qquad [\t_2(u),\t_2(v)]=0
}
(for the general simple Lie algebra ${\cal G}$ one should use the
corresponding Killing form).

The above quadratic invariant is enough to serve the $sl(2)$ case.
The definitions \ABtwo\ of $A(u)$ and $B(u)$ and, respectively, \Bxj\
of and \Axj\ of $x_j$ and $z_j$ remain the same as in the
quadratic case. The commutation relations \commzx\ and the quantum
characteristic equation \qchartwo\ are replaced, respectively, by
\eqn\commzxlin{
 [x_j,x_k]=[z_j,z_k]=0, \qquad [z_j,x_k]=z_j\d_{jk}
}
and
\eqn\qcharlin{
        z_j^2-\t_2(x_j)=0
}
(here we put $\eta=1$ for simplicity). Using then the realization
\eqn\relzlin{
  z_j=\dd_{x_j}+\zeta(x_j)
}
for $z_j$ one obtains for the separated spectral problem the second order
differential equation \SklGaudin\
\eqn\sepeqGaudin{
  \psi''+2\zeta\psi'+(\zeta^2+\zeta')\psi={\bf\tau}_2\psi.
}

It is easy to anticipate that for $GL(N)$ the separated equation
should become an $N$th order differential equation, though the
calculation still waits to be done.

For discussion of \SoV\ for the quadratic algebra \rLrL\
in the $sl(2)$ case (open Toda chain with boundary conditions) see
\KuzTsigBC.

The above scheme of quantum \SoV\ is general enough to serve the variety
of quantum integrable models obtained by taking concrete representations of the
yangian \RLL\ or the algebra \rLL. Below we shall illustrate on few
examples of $GL(2)$-type models
the diversity of possibilities arising. The main problem of adjusting the
general scheme to a concrete model is to find the spectrum of the
commuting operators $B(u)$ and, consequently, of $x_j$ and to describe the
functional space in which the one-dimensional spectral problem
\sepeqtwo\ or \sepeqGaudin\ has to be solved.

\newsec{XXX magnetic chain}
The general finite-dimensional irreducible representation of the yangian
${\cal Y}[sl(2)]$ is realized in the tensor product of $D$
finite-dimensional irreducible representations
of the Lie algebra $sl(2)$ (classically, $D$ is the number of degrees
of freedom)
\eqn\defSpins{\eqalign{
&  [S^3_m,S^\pm_n]=\pm S^\pm_n\d_{mn} \qquad
  [S^+_m,S^-_n]=2S^3_n\d_{mn} \qquad m,n=1,\ldots,D \cr
& (S_n^3)^2+\half(S^+_nS^-_n+S^-_nS^+_n)=l_n(l_n+1) \qquad
   l_n\in\{\half,1,\frac{3}{2},2,\ldots\}
}}
and can be written in the factorized form ({\it monodromy matrix})
\eqn\defLXXX{
  L_{\rm XXX}(u)
   =K{\cal L}_D(u-\d_D)\ldots {\cal L}_2(u-\d_2){\cal L}_1(u-\d_1)
}
where
\eqn\XXXLop{
  {\cal L}_n(u)=u\id+\eta S_n, \qquad
  S_n=\pmatrix{ S^3_n&S^-_n\cr S^+_n&-S^3_n},
}
and $K$ is a constant numeric matrix. It is customary to use the notation
$T(u)$ instead of $L(u)$ like e.g. in \SklNankai, but here we had to sacrifice
it to preserve the coherence of notation. The representation given by \defLXXX\
has dimension $\prod_{n=1}^D(2l_n+1)$ and is parametrized by: $2\times2$ matrix
$K$, number $D$ of degrees of freedom, spins $l_n$ and shifts $\d_n$.
The Casimir operator (quantum determinant \qtn) $t_2(u)$ takes the value
\eqn\qdetXXX{
t_2(u)=\det K\prod_{n=1}^D(u-\d_n-l_n\eta)(u-\d_n+l_n\eta+\eta).
}

The corresponding quantum integrable model is called {\it inhomogeneous
XXX magnet}.

The parameters of the representation being in the generic position,
the representation \defLXXX\ of the yangian turns out to be irreducible, and
the spectrum of the operators $x_j$ defined from \Bxj\ for $B(u)$ given by
\ABtwo\ turns out to be the finite set \SklNankai\
\eqn\specx{
 {\rm spec}[x_j]=\L_j
   =\{\l^-_j,\l^-_j+\eta,\ldots,\l^+_j-\eta,
   \l^+_j\}, \qquad \l^\pm_j=\delta_j\pm l_j\eta.
}

For the separated finite-difference equations \sepeqtwo\ to be well defined
on the finite set $\L_j$,
the coefficients $\D_\pm(x)$ must satisfy the boundary conditions
$$ \D_\pm(\d_n\pm l_n\eta)=0. $$

The most convenient choice of $\D_\pm(u)$ is
\eqn\defDpm{
 \D_\pm(u)=\kappa_\pm\prod_{n=1}^D(u-\d_n\mp l_n\eta), \qquad
   \kappa_+\kappa_-=\det K.
}

The $sl(2)$ Gaudin model is the degenerate case of XXX magnet obtained
in the limit $\eta\rightarrow 0$. The corresponding $L$ operator is produced
renormalizing $L$ operator \defLXXX, putting
$K:=1+\eta \K$, $\tr\K=0$, and expanding in $\eta$:
$$ \frac{L_{\rm XXX}(u)}{\prod_n(u-\d_n)}
   =1+\eta L_{\rm Gaudin}(u)+O(\eta^2).
$$

The $L$ operator
\eqn\defLGaudin{
  L_{\rm Gaudin}(u)=\K+\sum_{n=1}^D \frac{S_n}{u-\d_n}
}
satisfies the linear commutation relations \rLL\ with the $\r$
matrix \RGLN\ for $\eta=1$. The spectral invariant
$\t_2(u)$ \quadrinv\ produces the commuting hamiltonians which are
quadratic in spin operators
\eqn\Gaudint{
  \t_2(u)=\half\tr {\cal K}^2+\sum_{n=1}^D\frac{H_n}{u-\d_n}
           +\sum_{n=1}^D\frac{l_n(l_n+1)}{(u-\d_n)^2},
}
\eqn\GaudinHam{
  H_n=\tr \K S_n
   +\sum_{\scriptstyle m=1 \atop \scriptstyle  m\neq n}^D
      \frac{\tr S_mS_n}{\d_n-\d_m}.
}

The normalization \defDpm\ of $\D_\pm(u)$ in \sepeqtwo\ corresponds to
the normalization
\eqn\defzeta{
       \zeta(u)=c-\sum_{n=1}^D\frac{l_n}{u-\d_n}, \qquad c^2=-\det\K
}
of $\zeta(u)$ in \relzlin, \sepeqGaudin.

In the limit $\eta\rightarrow 0$ the $2l_j+1$ points of the spectrum
$\L_j$ of the operator $x_j$ merge into one point $x_j=\d_j$ of
multiplicity $2l_j+1$. The space of functions on $\L_j$ is
understood, respectively, as the
the ring of polynomials in $x_j$ factorized over the ideal
$(x_j-\d_j)^{2l_j+1}=0$. The spectrum of the hamiltonians \GaudinHam\
is given then by the values of $H_n$ in \Gaudint\ for which
the differential equation \sepeqGaudin\ with $\zeta(u)$ given by
\defzeta\ in each of the points $x=\d_j|_{j=1}^D$
has a regular solution
$\psi_j(x)=1+\sum_{k=1}^\infty (x-\d_j)^k\psi_j^{(k)}$, see \SklGaudin.

If one realizes the spin operators $S_n^\a$ \defSpins\ as the
differential operators
\eqn\reprspins{
  S_n^3=y_n\dd_{y_n}-l_n, \qquad
  S_n^+=y_n, \qquad
  S_n^-=y_n\dd^2_{y_n}-2l_n\dd_{y_n}
}
then the equation $B(x_j)=L_{21}(x_j)=0$ defines a ``purely
coordinate'', in the sense \cotan, change of variables
$\{y_n\}\rightarrow\{x_j\}$. The separated coordinates $x_j$ can be
described as generalized ellipsoidal coordinates, see \KuzGaudin.
In fact, all the models allowing \SoV\ in generalized ellipsoidal
coordinates, such as Neumann model \BabTal\ and its generalizations
\refs{\MacF,\HarWin} or Euler-Manakov top \KomKuzEuler,
can be considered as degenerate cases of Gaudin model
\KuzNeumann. For the generalization of Gaudin model to $osp(1|2)$ Lie
superalgebra see \BrMcf. An application of Gaudin model with $L(u)$
having a second order pole to the atomic physics (Coulomb three-body
problem) is considered in \KomCoulomb.

The particular simplicity of the Gaudin model makes it attractive
for rigorous mathematical analysis. Much attention was devoted
last years to the study of the connection between Gaudin model and
Knizhnik-Zamolodchikov (KZ) equations for the correlators in conformal
field theory. The fact that the eigenvalue problem for the Gaudin
hamiltonians \GaudinHam\ coincides with the KZ equations on the critical
level can be exploited both to produce integral representations
for the solutions to KZ equations \BabFlu\ and a new derivation of
the formula for the norm of the Gaudin eigenfunctions \ReshVar.
In the recent paper \FFR\ the representation theory for the affine Lie
algebras is applied to derive, in particular, Bethe equations for
the Gaudin model corresponding to arbitrary simple Lie algebra
and to reveal thus the algebraic meaning of Bethe ansatz.
Hopefully, the methods developed in \FFR\ will be useful also in
understanding the algebraic roots of \SoV.

The XXX model, as well as the $sl(2)$ Gaudin model,
presents a convenient possibility to compare the results
of \SoV\ method with those obtained by the Algebraic Bethe Ansatz
(ABA) \QISM.
Since the subject is discussed in detail in \refs{\SklFBA,\SklNankai,
\SklGaudin}
we present here only the summary of the analysis. The \SoV\ and ABA methods
lead to the same equation \sepeqtwo, resp.\ \sepeqGaudin,
though its interpretations differ.
In \SoV\ method the equation is solved on the finite set
$\L=\bigcup_j\L_j$
whereas in ABA $\psi(u)$ is supposed to be a polynomial whose zeroes
$v_m$ parametrize the Bethe vector $\Psi_{\vec v}$
\eqn\Bethevec{
 \Psi_{\vec v}=\prod_{m=1}^M L_{21}(v_m)\vac, \qquad
   L_{12}(u)\vac=0.
}

In the \SoV\ approach there is one-to-one correspondence between the
solutions of the problem \sepeqtwo\ on $\L$ and the eigenvectors of
the commuting quantum hamiltonians which is not the case in the ABA
approach where the so-called
``completeness problem'' arises. The \SoV\ method provides the basis for
the rigorous analysis of the completeness problem and allows to formulate
the criterion of completeness. In case of the Gaudin model the
criterion sounds as follows \SklGaudin. Let $Q(u)$ be a polynomial
solution to the differential equation \sepeqGaudin. If $Q(\d_j)\neq 0$
$\forall j=1,\ldots,D$ then the corresponding Bethe vector \Bethevec\
is nonzero. If, however, $Q(\d_j)=0$ for some $j$ then the
corresponding Bethe vector is zero, and the set of Bethe eigenvectors
is incomplete. Moreover, the nonzero eigenvector corresponding to such $Q(u)$
exists if and only if the linearly
independent solution of the differential equation is regular in the
same point $x=\d_j$.

The power of \SoV\ is revealed most obviously in the cases when the
representation of the yangian does not possess the highest vector
$\vac$ such that $L_{12}\vac=0$ and hence ABA cannot be applied.
These cases correspond to the infinite dimensional representations
of $sl(2)$ for the operators $S_n^\a$ \XXXLop.
The corresponding separated wave functions $\psi(x_j)$ are not
more polynomials, and the separated equations \sepeqtwo, \sepeqGaudin\
should be accompanied with some square integrability conditions.
Depending on the
real form of $sl(2)$ in question there is plenty of analytical
possibilities.

For the Goryachev-Chaplygin top the spectrum of $x_j$ is real and discrete,
and the shift $\eta$ in \sepeqtwo\ is real, see \refs{\KomGCh,\SklGoryachev}
and also \GCh\ for generalizations.

In case of the Toda lattice the spectrum of $x_j$ is real and the
shifts $\eta$ in \sepeqtwo\ are imaginary \SklToda. See also
\KuzTsigToda\ for the relativistic version and \KuzTsigBC\ for the
lattice with boundary conditions.

A version of noncompact XXX magnet  applied recently \QCD\
to describe the QCD in the asymptotic high energy regime
also does not have the highest weight vector, so the \SoV\ is
the natural approach to try.

For the analogous effects in the Neumann model, see \BabTal.

\newsec{Infinite volume limit}
So far, we discussed only the integrable models with a finite number
$D$ of degrees of freedom. The passage to $D=\infty$ can be made in
two ways: either by taking the continuum limit or the infinite volume
limit. In the continuum limit the representation \defLXXX\ of the
monodromy matrix $L(u)$ as a product of local ${\cal L}$ operators
${\cal L}_n(u)$
is replaced by the representation in the form of the ordered
exponential \refs{\QISM,\SklNLS}
\eqn\ordexp{
L_{\xi_-}^{\xi_+}(u)=
:\overrightarrow{\exp}\int_{\xi_-}^{\xi_+}\LL(u,\xi)\,d \xi:.
}

For example, for the nonlinear Schr\"odinger equation, described
in terms of the canonical fields $\Psi(\xi)$, $\Psi^*(\xi)$
\eqn\commPsi{
  [\Psi(\xi),\Psi(\xi')]=[\Psi^*(\xi),\Psi^*(\xi')]=0, \qquad
    [\Psi(\xi),\Psi^*(\xi')]=\d(\xi-\xi')
}
acting in the Fock space $\Psi(\xi)\vac=0$, the infinitesimal
$L$ operator $\LL(u,\xi)$ is given by
\eqn\LNLS{
    \LL(u,x)=\pmatrix{
           -i u/2 & \sqrt c\,\Psi^*(\xi)\cr
  \sqrt c\,\Psi(\xi) & {i}u/2
}}

The corresponding quantum monodromy matrix $L_{\xi_-}^{\xi_+}(u)$
given by \ordexp, where $:\;:$ stands for the normal ordering,
satisfies the commutation relations \RLL\ with the $R$ matrix
of XXX type \RGLN\ and $\eta=-ic$. The quantum determinant $t_2(u)$
of $L_{\xi_-}^{\xi_+}(u)$ is equal to $e^{-cV/2}$, and the trace $t_1(u)$
of $L_{\xi_-}^{\xi_+}(u)$ generates the commuting hamiltonians,
in particular
\eqn\HamNLS{
H=\int_{\xi_-}^{\xi+}(\Psi_\xi^*\Psi_\xi^{\phantom{*}}+
     c\Psi^*\Psi^*\Psi\Psi)\,d \xi.
}

We assume that the fields $\Psi(\xi)$, $\Psi^*(\xi)$ are periodic in
$\xi$ with the period $V=\xi_+-\xi_-$ and the coupling constant $c$
is positive $c>0$ (repulsive case).

It is convenient to choose the following normalization $\vec \a(u)$ of
the Baker-Akhiezer function \normOm
$$ \a_1(u)=1,\qquad \a_2(u)=i. $$

The reason for such a choice is that
the corresponding operator family $B(u)$ \defBaa\
\eqn\defBNLS{
    B(u)=-iL_{11}(u)+L_{12}(u)+L_{21}(u)+iL_{22}(u)
}
has the symmetry  $B(u)^*=B(\bar u)$ and its zeroes $x_j$ are self-adjoint
operators.

It is easy to find that the separated equation \sepeqtwo\ takes the
form
\eqn\sepeqNLS{
 \tau_1(x)\psi(x)=e^{-ixV/2}\psi(x+ic)+e^{ixV/2}\psi(x-ic).
}

Since $B(u)$, like $L_{\xi_-}^{\xi_+}(u)$, is not a
polynomial  but a holomorphic function of $u\in\CC$, it has an infinite
discrete set
of zeroes $x_j$ having asymptotics $x_j=2\pi j/V+O(j^{-1})$ as
$j\rightarrow\infty$. The necessity to handle the functions of infinite
number of variables complicates greatly the justification of the
standard \SoV\ construction even in the classical case.

The situation, however, simplifies drastically in the infinite
volume limit (we consider the zero density case described again by the
Fock representation for $\Psi(\xi)$, $\Psi^*(\xi)$).
The definition of the monodromy matrix $L(u)$ for $V=\infty$ needs
some caution. If one tries to pass to the limit
$\xi_\pm\rightarrow\pm\infty$ directly in the expression \ordexp\
one finds immediately that, in order to get a finite expression,
it is necessary to cancel the asymptotic
behaviour of $L_{\xi_-}^{\xi_+}(u)$ introducing the exponential factors
\eqn\defbadL{
  L_{-\infty}^{+\infty}(u)=\lim_{\xi_\pm\rightarrow\pm\infty}
   e^{i u\s_3\xi_+/2}
   L_{\xi_-}^{\xi_+}(u)
   e^{-iu\s_3\xi_-/2},
   \qquad  \s_3=\pmatrix{1 & 0 \cr 0 & -1}
}
which, however, destroy completely the nice commutation relations
\RLL. The solution of the problem is to factorize first the
finite-volume monodromy matrix
$L_{\xi_-}^{\xi_+}(u)=L_{\xi_0}^{\xi_+}(u)L_{\xi_-}^{\xi_0}(u)$
and to permute the factors:
$L_{\xi_0}^{\xi_0}(u)=L_{\xi_-}^{\xi_0}(u)L_{\xi_0}^{\xi_+}(u)$.
This operation does not change the quantity $t_1(u)=\tr L(u)$
generating the integrals of motion. The matrix $L_{\xi_0}^{\xi_0}(u)$,
however, behaves in the limit $\xi_\pm\rightarrow\pm\infty$ much better:
$$ L_{\xi_0}^{\xi_0}(u)
      =L^{\xi_0}_{-\infty}e^{-iuV\s_3/2}L^{+\infty}_{\xi_0}
      \sim e^{\mp iuV/2}L^{\xi_0}_{-\infty}P_\pm L^{+\infty}_{\xi_0}
$$
where
$$P_+=\pmatrix{
 1&0\cr 0&0},\qquad
  P_-=\pmatrix{0&0\cr0&1},
$$
and the upper (lower) sign corresponds, respectively, to
${\rm Im\,}u>0$\ $(<0)$.

The scalar factor $e^{\mp iuV/2}$ can be cancelled out of
$L_{\xi_0}^{\xi_0}(u)$ since it does not affect the relation \RLL.
Hence, we can take for the monodromy matrix in the infinite volume
the matrix
\eqn\defLgood{
      L(u)=L^{\xi_0}_{-\infty}P_\pm L^{+\infty}_{\xi_0}.
}

The matrix $L(u)$ is analytical in the complex plane of $u$ except
the cut along the real axis. Its quantum determinant $t_2(u)$ is zero.

It is interesting to compare the above results with those of Kyoto
group on the XXZ magnetic chain \refs{\Miwa}. In both cases the
infinite volume case is considered, and a sort of monodromy
matrix $L(u)$ satisfying \RLL\ is constructed. In the XXZ case, however,
all components of  $L(u)$ are integrals of motion whereas in the
nonlinear Schr\"odinger case only $\tr L(u)$ is one. Moreover, in the
XXZ case $L(u)$ is analytic in $\CC$ whereas in the NLS case $L(u)$
has a cut along the real axis. The above distinctions are probably due
to the different nature of the vacuum state: antiferromagnetic for
XXZ and ferromagnetic for NLS which in the last case requires
introducing the asymptotic exponents $e^{\pm iu\s_3\xi_\pm/2}$.

Let us consider now how the whole \SoV\ construction is modified in the
infinite volume limit.
As  $\xi_\pm\rightarrow\pm\infty$ the zeroes $x_j$
of $B(u)$ accumulate to a continuous distribution
$$ \frac{2\pi}{V}j\rightarrow \nu, \qquad
   x_j\sim \nu+\frac{2\pi}{V}q(\nu)
$$
with the density $q(\nu)$. The Hilbert space ${\cal H}$ of quantum states is
realized then as the space of the linear functionals $W[q(\nu)]$
of $q(\nu)$, $\nu\in\RR$ which are square integrable
$$   \|W\|^2=\int |W[q(\nu)]|^2 \,\d m.
$$
with respect to the measure $\d m$ which, fortunately, happens
to be Gaussian and is characterized uniquely by the correlator
(covariance kernel)
$$ \left<q(\mu)q(\nu)\right>_\nu=
  \frac{1}{4\pi^2}\ln\left(1+\frac{c^2}{(\mu-\nu)^2}\right).$$

The representation of $B(u)$ as an infinite
product $B(u)\sim\prod_j (u-x_j)$ is replaced by the Cauchy integral
\eqn\Cauchy{
B(u)=\mp\frac{i}{2}\exp\left\{-\int_{-\infty}^{+\infty}\frac{d \nu}{u-\nu}
q(\nu)\right\}
}
where the upper (lower) sign corresponds, respectively, to
${\rm Im\,}u>0$\ $(<0)$.

Though the quantity $t_1(u)$ is represented in ${\cal H}$ as a
complicated variational operator, its eigenfunctions nevertheless
can be found exactly and have
rather simple structure which is natural to consider as a continual
analog of \SoV. The separated equation \sepeqNLS\ is replaced
respectively by a boundary problem for analytical functions having
a cut along the real axis. Unfortunately, the complications due to
the cut make a brief explanation impossible and for the details we
refer the reader to the original papers \SklNLSiv.

For the nonlinear Schr\"odinger equation the above \SoV\ procedure
can be justified rigorously by comparison with the results known
from the algebraic Bethe Ansatz \SklNLSiv. The analogous construction
for the relativistic sinh-Gordon model, though not so
rigorous, can also be performed \SklSinhG\ and leads to quite
reasonable results concerning the spectrum of the model. It would be
interesting to generalize the results of \SklSinhG\ to the relativistic
Toda field theories.

\newsec{Classical XYZ magnet}
The XYZ magnet provides an example of quite nontrivial normalization
of the Baker-Akhiezer function $\Omega(u)$ necessary to produce \SoV.
The construction presented below is taken from \SklXYZ.
The model is described in terms of the $2\times 2$ matrix $L(u)$
satisfying the relation \pblu\ with the $r$-matrix
\eqn\rXYZ{
  r(u)=\sum_{\a=1}^3 w_\a(u)\one\s_{\!\a}\two\s_{\!\a}
}
where $\s_\a$ are standard Pauli matrices
\eqn\Pauli{
 \s_1=\pmatrix{0 & 1 \cr 1 & 0}, \qquad
 \s_2=\pmatrix{0 & -i \cr i & 0}, \qquad
 \s_1=\pmatrix{1 & 0 \cr 0 & -1},
}
and $w_\a(u)$ are certain elliptic functions whose exact expression
is not important for the moment. It suffices to remark that $r(u)$ is
a meromorphic function on $\CC$ having simple poles on the periodic
lattice
$\Gamma=\{u\in\CC | u=m+\tau n;\quad m,n\in\ZZ;\quad {\rm Im}\,\tau>0\}$
and possessing the periodicity properties
\eqn\periodr{\eqalign{
  r(u+1)&=\one\s_1 r(u)\one\s_1=\two\s_1 r(u)\two\s_1, \cr
  r(u+\tau)&=\one\s_3 r(u)\one\s_3=\two\s_3 r(u)\two\s_3.
}}

The $L$ operator $L(u)$, in turn, is a holomorphic function
characterized by the quasiperiodicity properties
\eqn\periodL{
  L(u+1)=(-1)^D\s_1 L(u)\s_1, \qquad
  L(u+\tau)=(-1)^De^{-i\pi D(u+\tau)}\s_3 L(u)\s_3
}
where $D$ is a positive integer. The conditions \periodL\ determine
$L(u)$ up to 4D free parameters \refs{\Mumford}
which can be considered as the
dynamical variables with the Poisson structure defined by \rLL.
Since, however, the determinant $\det L(u)$ generating the center of the
Poisson algebra contains $2D$ parameters (Casimir functions), their
values can be fixed which leaves $2D$-dimensional phase space, the
constant $D$ being thus the number of degrees of freedom.

In contrast with the XXX magnet, the normalization $\vec\a(u)={\rm const}$
of $\Omega(u)$
does not produce \SoV\ for the XYZ magnet for any $\vec\a$. The reason
is that the corresponding function $B(u)$ has no definite quasiperiodicity
for the period lattice $\Gamma$ and can be characterized only
by the periodicity properties in $2\Gamma$
$$ B(u+2)=B(u), \qquad B(u+2\tau)=e^{-i\pi D(2u+3\tau)}B(u)  $$
from which it follows that $B(u)$ has $4D$ zeroes in the
fundamental region $\CC/2\Gamma$ whereas one needs only $D$ separated
coordinates $x_j$.

The correct normalization \SklXYZ\
is given by the holomorphic functions $\vec\a(u)$
having the periodicity properties
\eqn\defalpha{\eqalign{
  \a_1(u+1)&=\a_1(u), \qquad
  \a_1(u+\tau)=-e^{-i\pi(u+\tau-y)/2}\a_1(u),  \cr
  \a_2(u+1)&=-\a_2(u), \qquad
  \a_2(u+\tau)=e^{-i\pi(u+\tau-y)/2}\a_1(u)
}}
where $y$ is a parameter. The explicit expressions for $\a_n(u)$ can be
given in terms of theta-functions for the lattice of periods $\Gamma$
\SklXYZ.
The function $B(u)$ corresponding to the normalization vector
\defalpha\ is given by the formula \defBaa\ and has good periodicity
properties on the lattice $\Gamma$
\eqn\periodB{
 B(u+1)=(-1)^{D+1}B(u), \qquad
 B(u+\tau)=(-1)^{D+1}e^{-i\pi (D+1)(u+\tau)}e^{-i\pi y}B(u).
}

Consequently, $B(u)$ is a theta-function of order $D+1$ and has $D+1$
zeros in the fundamental region $\CC/\Gamma$. It remains to require
one superficial zero of $B(u)$ to be a constant ($c$-number)
\eqn\fixy{
  B(\xi)=0
}
which can be considered as the equation determining the parameter $y$
as a function on the phase space. The remaining  $D$ zeroes of $B(u)$
are candidates for the separated coordinates $x_j$. A direct, though
cumbersome,
calculation of the Poisson brackets \SklXYZ\ validates successfully
the conjecture.

An open question is the quantization of the above construction.
The difficulty is the ordering problem in the expression \defBaa\
since the quantities $\a_n(u)$ contain dynamical variables through
their dependence on $y$.

The example of XYZ model raises the question if the correct
normalization producing \SoV\ could be found for other integrable models
where the simplest choice $\vec\a(u)={\rm const}$ is known to fail
\refs{\nonSLN}.

\newsec{3-particle Calogero-Moser model}

In this section we present new results for
the classical and quantum 3-particle elliptic Calogero-Moser model.
The model is interesting as an example of \SoV\ in case of a dynamical
r-matrix.

In the classical case the model has 3 degrees of freedom and
is described in terms of the canonical variables $(\pi_n,q_n)$
\eqn\pbCM{
 \{q_m,q_n\}=\{\pi_m,\pi_n\}=0,\qquad  \{\pi_m,q_n\}=\d_{mn},
 \qquad m,n=1,2,3.
}

The commuting hamiltonians are \OlPer\
\eqn\HamCM{\eqalign{
 H_1 & =  \pi_1+\pi_2+\pi_3 \cr
 H_2 & =  \pi_1\pi_2+\pi_1\pi_3+\pi_2\pi_3
      -g^2(\wp(q_{12})+\wp(q_{13})+\wp(q_{23})) \cr
 H_3 & =  \pi_1\pi_2\pi_3
       -g^2(\pi_1\wp(q_{23})+\pi_2\wp(q_{13})+\pi_3\wp(q_{12}))
}}
where $q_{mn}=q_m-q_n$, $\wp$ is Weierstrass elliptic function,
and $g$ is the coupling constant. The corresponding $L$ operator is
\eqn\defLCM{
    L_{\rm CM}(u)=\pmatrix{
        \pi_1 & -igQ_{12}(u) & -igQ_{13}(u) \cr
     -igQ_{21}(u) & \pi_2 & -igQ_{23}(u) \cr
     -igQ_{31}(u) & -igQ_{32}(u) & \pi_3 }, \quad
}
where
\eqn\defQ{
    Q_{mn}=\frac{\s(u+q_{mn})}{\s(u)\s(q_{mn})},
}
and $\s(u)$ is Weierstrass sigma function.

The hamiltonians \HamCM\ can be obtained from the spectral invariants of
the $L$ operator
\eqn\charpolCM{
 \det(z-L(u))=z^3-z^2t_1(u)+zt_2(u)-t_3(u)
}
\eqn\deftCM{\eqalign{
 t_1(u) & =  H_1 \cr
 t_2(u) & =  H_2+3g^2\wp(u) \cr
 t_3(u) & =  H_3+g^2\wp(u)H_1-ig^3\wp^\prime(u)
}}

The $L$ operator \defLCM\ satisfies the identity \pbl\ with rather
complicated $r$-matrix depending on $q$ \rmCM. In the absence
of general theory of dynamical $r$-matrices the only available
strategy is to try one-by-one possible ans\"atze for the normalization
$\vec\a(u)$ of the Baker-Akhiezer function $\Omega(u)$. Fortunately,
the very first attempt succeeds: the simplest normalization
\normGLN\ which was applied to the $GL(N)$-magnet does also produce
\SoV\ for the Calogero-Moser model.

For our purposes it is convenient to write down the set of equations
for the pair $(x,z)$ as \eqAA\
where the functions $A_{1,2}(u)$ are given by the formulas \Athree\
and, in our case, are
\eqn\defACM{\eqalign{
 A_1(u)&=\pi_1+ig{[}\zeta(u)-\zeta(u-q_{23})+\zeta(q_{12})-\zeta(q_{13}){]}\cr
 A_2(u)&=\pi_2+ig{[}\zeta(u)-\zeta(u-q_{13})-\zeta(q_{12})-\zeta(q_{23}){]}
}}
where $\zeta(u)$ is Weierstrass zeta function.

Since the $r$-matrix for the CM model is different from \rgln\
we cannot rely on the results obtained for the $GL(N)$ magnet and have
to calculate the Poisson brackets between $z$ and $x$ directly.
It turns out that the equations \eqAA\ have only two solutions:
$(z_1,x_1)$ and $(z_2,x_2)$.
For the third pair of variables one has to take $(P,Q)$
\eqn\defPQ{
  P=\pi_1+\pi_2+\pi_3, \qquad Q=q_3.
}

The calculation shows that the variables $(P,z_1,z_2;Q,x_1,x_2)$
are canonical and satisfy the relations
\eqn\sepeqCM{\eqalign{
&   P-H_1=0 \cr
&  z_j^3-z_j^2H_1+z_j(H_2+3g^2\wp(x_j))
  -(H_3+g^2\wp(x_j)H_1-ig^3\wp^\prime(x_j))=0
}}
which fit the form \sovcl\ and provide thus a \SoV.

In the quantum case the momenta $\pi_j$ are realized as the differentiations
$\pi_j=-i\dd_{q_j}$. The hamiltonians \HamCM,
respectively, are replaced by the differential operators
\eqn\HamCMq{\eqalign{
 H_1&=-i(\dd_{q_1}+\dd_{q_2}+\dd_{q_3}), \cr
 H_2&=-\dd^2_{q_1q_2}-\dd^2_{q_1q_3}-\dd^2_{q_2q_3}
   -g(g-1){[}\wp(q_{12})+\wp(q_{13})+\wp(q_{23}){]}, \cr
 H_3&=i\dd^3_{q_1q_2q_3}+ig(g-1)
    {[}\wp(q_{23})\dd_{q_1}+\wp(q_{13})\dd_{q_2}+\wp(q_{12})\dd_{q_3}{]}
}}
which do commute \OlPer, as their classical counterparts.

Since it is still unknown how to quantize the relation \pbl\
in case of the dynamical $r$-matrices, we again have to rely on good
luck trying to find a quantum \SoV. An additional obstacle is provided
by the fact that in our case, even classically, $\{B(u),B(v)\}\neq0$.
Therefore, there is little hope to construct quantum $x_j$ as zeroes
of a commuting family of operators $B(u)$ like
in case of the $GL(N)$ magnet. Instead, we shall rather look for the
kernel $K(x_1,x_2,Q|q_1,q_2,q_3)$ of the integral operator
(classically, canonical transformation) intertwining the $xQ$ and
$q$ representations.

The first of the classical separated equations \sepeqCM\ is easy
to quantize. It expresses the conservation of the total momentum
and allows to eliminate one pair of variables from $K$:
$$ \left\{\eqalign{
    P&=-i(\dd_{q_1}+\dd_{q_2}+\dd_{q_3})=-i\dd_Q \cr
    Q&=q_3}\right.
   \quad \Longrightarrow \quad
   \left\{\eqalign{
     (\dd_{q_1}+\dd_{q_2}+\dd_{q_3}+\dd_Q)K&=0 \cr
     (Q-q_3)K&=0}\right.
$$
$$ (Q-q_3)K=0 \quad\Longrightarrow\quad
    K=\d(Q-q_3)\tilde{K}(x_1,x_2|q_1,q_2,q_3), $$
$$\eqalign{
     (\dd_{q_1}+\dd_{q_2}+\dd_{q_3}+\dd_Q)K=0 &\Longrightarrow
          (\dd_{q_1}+\dd_{q_2}+\dd_{q_3})\tilde{K}=0, \cr
      &\Longrightarrow \tilde{K}=\tilde{K}(x_1,x_2|q_{13},q_{23}).}
$$

To determine the kernel $\tilde K$, let us try to quantize the equations
\defACM. Making the substitutions
$$ \pi_j\rightarrow i\dd_{q_{j3}}, \qquad
   z_j\rightarrow -i\dd_{x_j}, \qquad
   g\rightarrow g-1
$$
(the last one is a quantum correction found experimentally)
one obtains from \defACM\ the system of 4 first order differential
equations for $\tilde K$
\eqn\eqKtilde{\eqalign{
 (\dd_{x_1}+\dd_{q_{13}})\tilde{K}
+(g-1)[\zeta(x_1)-\zeta(x_1-q_{23})-\zeta(q_{13})+
\zeta(q_{13}-q_{23})]\tilde{K}&=0 \cr
 (\dd_{x_2}+\dd_{q_{13}})\tilde{K}
+(g-1)[\zeta(x_2)-\zeta(x_2-q_{23})-\zeta(q_{13})+
\zeta(q_{13}-q_{23})]\tilde{K}&=0 \cr
 (\dd_{x_1}+\dd_{q_{23}})\tilde{K}
+(g-1)[\zeta(x_1)-\zeta(x_1-q_{13})-\zeta(q_{23})
-\zeta(q_{13}-q_{23})]\tilde{K}&=0 \cr
 (\dd_{x_2}+\dd_{q_{23}})\tilde{K}
+(g-1)[\zeta(x_2)-\zeta(x_2-q_{13})-\zeta(q_{23})
-\zeta(q_{13}-q_{23})]\tilde{K}&=0
}}
which are easily solved producing the result which most
conveniently can be written using the variables $x_\pm=x_1\pm x_2$ and
$\xi_\pm=q_{13}\pm q_{23}$:
$$
\tilde{K}=\d(x_+-\xi_+){\cal K}(x_+,x_-;\xi_-)
$$
\eqn\defKK{
{\cal K}=\left[\frac{
   \s\left(\frac{\displaystyle \xi_-+x_-}{\displaystyle 2}\right)
   \s\left(\frac{\displaystyle \xi_--x_-}{\displaystyle 2}\right)
   \s\left(\frac{\displaystyle x_++\xi_-}{\displaystyle 2}\right)
   \s\left(\frac{\displaystyle x_+-\xi_-}{\displaystyle 2}\right)}{
 \s(x_1)\s(x_2)\s(\xi_-)}\right]^{\displaystyle g-1}.
}

The above argument, of course, has only heuristic value and provides
no guarantee that the kernel $K$ thus constructed would produce \SoV.
What is neccessary to verify is that the integral operator
with the kernel $K$ transforms an eigenfunction $\Psi(q_1,q_2,q_3)$
of the hamiltonians $H_n$ \HamCM\ satisfying \HPsi\
into the  function $\tilde \Psi(x_1,x_2,Q)$ satisfying separated
equations of the type \sepeqPsi.

The observation which is crucial for establishing \SoV\ is that
the kernel $K$ satisfies the differential equations
\edef\diffeqK{\onum}
$$\eqalignno{[-i\dd_Q-H_1^*]K&=0, &(\num{\rm a}) \cr
 \Bigl[i\dd^3_{x_j}+H_1^*\dd^2_{x_j}
   -i(H_2^*+3g(g-1)\wp(x_j))\dd_{x_j} \qquad\qquad\qquad &\cr
 -(H_3^*+g(g-1)H_1^*\wp(x_j)-ig(g-1)(g-2)\wp^\prime(x_j))\Bigr]K&=0
 &(\num{\rm b})
}$$
\advance\meqno by1
where $H_n^*$ is the Lagrange adjoint of $H_n$
$$ \int\phi(q)(H\psi)(q)\,dq=\int(H^*\phi)(q)\psi(q)\,dq. $$

The equations \diffeqK\ can be interpreted as the quantum analog
of the equations \sepeqCM\ (note the quantum corrections in $g$!).
Consider now the integral transform
\eqn\inttrbad{
\tilde\Psi(x_1,x_2,Q)=\int\!\!\!\int\!\!\!\int dq_1dq_2dq_3\,
  K(x_1,x_2,Q|q_1,q_2,q_3)\Psi(q_1,q_2,q_3)
}

Acting on $\tilde\Psi$ with the differential operators

$$\eqalignno{{\cal Q}&=-i\dd_Q-h_1, &(\num{\rm a}) \cr
{\cal D}_j&= i\dd^3_{x_j}+h_1\dd^2_{x_j}
   -i(h_2+3g(g-1)\wp(x_j))\dd_{x_j}  \cr
&\quad -(h_3+g(g-1)h_1\wp(x_j)-ig(g-1)(g-2)\wp^\prime(x_j))
&(\num{\rm b})
}$$
\advance\meqno by1
and supposing that $\Psi(\vec q)$ is an eigenfunction of $H_n$,
perform the integration by parts using the relations
\diffeqK. The resulting bulk part of the integral is zero. It remains
only to find such the limits of integration which would not contribute
to the result. Omitting the details of the guesswork we report only
the final result.

The function $\tilde\Psi(x_+,x_-,Q)$ obtained from an eigenfunction
$\Psi(\xi_+,\xi_-,q_3)$ of $H_n$ via the integral transform
\eqn\finalinttr{
  \tilde\Psi(x_+,x_-,Q)=\int_{x_-}^{x_+}d\xi_-\,{\cal K}(x_+,x_-;\xi_-)
  \Psi(x_+,\xi_-,Q)
}
satisfies the differential equations
$$ {\cal Q}\tilde\Psi=0, \qquad
    {\cal D}_j\tilde\Psi=0 $$
which imply the \SoV\
$$ \tilde{\Psi}(x_1,x_2,Q)=e^{ih_1Q}\psi(x_1)\psi(x_2) $$
where $\psi(x)$ satisfies a third-order analog of the Lam\'e
differential equation
\eqn\sepeqCMpsi{\eqalign{
& i\psi^{\prime\prime\prime}+h_1\psi^{\prime\prime}
  -i(h_2+3g(g-1)\wp(x))\psi^\prime \cr
&\qquad  -(h_3+g(g-1)\wp(x)h_1-ig(g-1)(g-2)\wp^\prime(x))\psi=0.
}}

The question of the correct boundary conditions for the separated
equation \sepeqCMpsi\ is presently under study.

For the degenerate case of trigonometric potential which corresponds
to replacing $\wp(q)$ by $1/\sin^2q$ in \HamCM\ and $\s(u)$ by
$\sin u$ in \defQ\ the spectrum and eigenfunctions of $H_n$ have been
well known quite a while ago \OlPer. The eigenfunctions are labelled by
triplets $\vec\nu=(\nu_1,\nu_2,\nu_3)$ of integers $\nu_j$ such that
$\nu_1\leq\nu_2\leq\nu_3$. The corresponding eigenvalues of $H_n$
are given by
\eqn\specHCM{
 \left\{\eqalign{
       h_1&=2(\mu_1+\mu_2+\mu_3) \cr
       h_2&=4(\mu_1\mu_2+\mu_1\mu_3+\mu_2\mu_3) \cr
       h_3&=8\mu_1\mu_2\mu_3
       }\right. \qquad
   \left\{\eqalign{
      \mu_1&=\nu_1-g \cr
      \mu_2&=\nu_2 \cr
      \mu_3&=\nu_3+g
     }\right.
}

The eigenfunctions $\Psi_{\vec\nu}$ have the structure
$\Psi_{\vec\nu}(\vec q)=\Psi_{000}(\vec q)\Phi_{\vec\nu}(\vec q)$ where
$\Psi_{000}(\vec q)=\sin^g q_{12}\sin^g q_{13}\sin^g q_{23}$ is
the vacuum eigenfunction corresponding to $\vec\nu=(0,0,0)$ and
$\Phi_{\vec\nu}(\vec q)$ are symmetric Laurent polynomials in variables
$t_j=e^{2iq_j}$ known as {\it Jack polynomials} \Macd.

The \SoV\ is produced by the same integral kernel $K$ up to
replacing $\s$ with $\sin$ in \defKK. The separated eigenfunctions
$\psi(y)$ satisfy the differential equation
\eqn\sepeqntrig{
 i\psi^{\prime\prime\prime}+h_1\psi^{\prime\prime}
-i\left(h_2+3\frac{g(g-1)}{\sin^2x}\right)\psi^\prime
  -\left(h_3+\frac{g(g-1)}{\sin^2x}h_1
    +2ig(g-1)(g-2)\frac{\cos x}{\sin^3 x}\right)\psi=0
}
and can be factorized $\psi_{\vec\nu}(x)=\psi_{000}(x)\phi_{\vec\nu}(x)$
into the product of the vacuum factor $\psi_{000}(x)=\sin^{2g}x$ and a Laurent
polynomial $\phi_{\vec\nu}$ in variable $t=e^{2ix}$. Despite the huge
body of facts known about the Jack polynomials, the last factorization
property seems to be a new result.

A more detailed exposition of the above results will be published
elsewhere \SklKuz.

\newsec{Discussion}
The above examples show the diversity of models allowing \SoV\
and give a support to the opinion that the domain of \SoV\ method
might be very large, even including all the models subject to the
classical Inverse Scattering Method and their quantum counterparts.

Let us enlist, in conclusion, some problems whose solution could
strengthen the positions of \SoV.
The most obvious object of study is provided by the class of integrable
models described by numeric unitary $R$-matrices, and in the first turn,
$sl(n)$-invariant magnets. The $sl(2)$ case being well enough studied,
the $sl(N)$ case seems to present only calculational difficulties.
The case of trigonometric $R$-matrices should not differ considerably
from the rational ($sl(N)$-invariant) case. For instance, it would be
interesting to generalise the results of section 5 to the relativistic
Toda field theories. In the case of elliptic
$R$-matrices discussed in section 6 the problem of notrivial normalization
of the Baker-Akhiezer function arises which leads to the complications
with the quantization. However, the success with the classical XYZ magnet
allows to hope that the quantization problem would be solved.
A similar, but more difficult problem of choosing the correct normalization
of B-A function arises in case of $R$-matrices corresponding to the simple
Lie algebras other than  $SL(N)$. The problem is not yet solved even in
the classical case. The most difficult problems arise in case of dynamical
$r$-matrices where neither general theory exists, no quantization rules are
known.

The problem which may be more important than studying all the particular
examples is to understand the algebraic structures underlying the \SoV\
and to explain why the ``magic recipe'' of taking the poles of B-A function
does work. The recent paper \FFR\ might be the first step in that direction.

\listrefs
\end